\documentclass[seceqn,dvips]{arxstspdf}
\usepackage{flushend}

\volume{24}
\issue{3}
\pubyear{2009}
\firstpage{280}
\lastpage{293}
\doi{10.1214/09-STS277C} \referstodoi{10.1214/09-STS277}

\makeatletter
\newtheorem{theorem}{Theorem}
\makeatother

\begin{document}
\begin{frontmatter}

\vspace*{6pt}

\title{Decoding the H-likelihood}
\runtitle{Discussion}
\pdftitle{Discussion on Likelihood Inference for Models with
Unobservable: Another View by Y. Lee and A. Nelder}

\begin{aug}
\author[a]{\fnms{Xiao-Li} \snm{Meng}\ead[label=e1]{meng@stat.harvard.edu}}
\runauthor{X.-L. Meng}

\affiliation{Harvard University}

\address[a]{Xiao-Li Meng is Whipple V. N. Jones Professor and Chair of Statistics,
Department of Statistics, Harvard University,
Cambridge, Massachusetts, USA \printead{e1}.}

\end{aug}


%

\end{frontmatter}
%

\section{Prologue}\label{sec1}
The~invitation for this discussion contribution came at the busiest
time in my (professional) life with four courses and many more meetings
attempting to compensate, psychologically, for the lost endowment at
Harvard. I could not possibly, however, decline David Madigan's kind
invitation. The~topic is dear to my heart, as it should be to any
statistician's, for without ``unobservables,'' we would be unemployable.
And I always wanted to know what ``h-likelihood'' is! I first heard the
term from my academic twin brother, Andrew Gelman, who sent me his
discussion of Lee and Nelder (\citeyear{LN1996}). Gelman's conclusion was that ``To
the extent that the methods in this paper give different answers from
the full Bayesian treatment, I would trust the latter.'' This of course
did not entice me to read the paper. Indeed, I still did not know its
definition when
I started to type this Prologue, nor have I had any professional or
personal contact with either author. I surmise this qualifies me as an
\textit{objective} discussant, though I hope in this case the phrase
\textit{objective} is not exchangeable with \textit{noninformative} or
\textit{ignorant}!

But surely, one may quibble, Gelman's comment must have influenced me.
True, but I'm not the kind of Bayesian who is unwilling to change
his/her prior. My pure interest is to decode the h-likelihood. If my
brother is right, I'll be more proud of him. If he is wrong, I'll be
wiser by learning something new. (But I do ask Professors Lee and
Nelder for their tolerance as I try to follow my brother's critical
style, in the name of good discussion!) So here I am, setting aside the
72-hour Memorial Day weekend, after persuading my teenagers that their
father's H-bomb mystery is more urgent to solve than his colleague Dr.
Langdon's prevention of the antimatter explosion in 24 hours, which
actually repeats every weekend.

%
\section{Preparing for a Bayesian Inference of~H-likelihood}\label{sec2}

\subsection{Prior Formulation}
Naturally I will adopt a Bayesian approach to infer what is the real
``H'' in the h-likelihood. What could it actually stand for? (I)
Heuristic argument? (II)~Handy approximation? (III) Hybrid method? Or
even (IV)~Hidden treasure? Of course, a priori I would not be a good
Bayesian if I exclude ``(V) Hype?'' no matter how small my prior belief
in it. Gelman's comment led me to assign the highest prior probability
to (III), 60\%. Since the events here are clearly not mutually
exclusive, (I) and (II) also deserve some nontrivial prior
probabilities which are 40\% each for reasons I can only explain to
myself. But for reasons everyone can explain, the prior probabilities
for categories such as (IV) or (V) are best kept confidential, other
than that they of course depend on one's knowledge of the author(s) and
the journal.

\subsection{Data Collection}
Immediately, I ran into the usual problem of any real-life data
collection---there are never enough time or resources! It is already
2:31 pm Saturday as I am typing this sentence, and I yet need to read
the paper plus four reference papers I was able to download from \mbox{JSTOR}:
two discussion papers by the same authors (Lee and Nelder, \citeyear{LN1996},
\citeyear{LN2004})
and the two papers in \textit{Biometrika} that illustrate the use of
h-likelihood (Ha, Lee and Song, \citeyear{HLS2001}, Lee and Nelder, \citeyear{LN2001}). Lee,
Nelder and Pawitan's (\citeyear{LNP2006}) book of course would be invaluable which,
unfortunately, turns out to be literally true in this case because
apparently no Harvard library can afford it.

So I settle with these four papers as background, knowing well the
potential bias due to my haphazard selection and all the
``unobservables'' to me at this moment. Hence my apologies to the
authors---and readers---in advance. To compensate for my hastiness,
I'll actually read all five papers, and the discussions, before forming
my likelihood, with or without ``H''!

%
\subsection{Data Processing}
Another grand challenge in real-life statistical analysis is \textit
{data processing}, something that unfortunately has not received nearly
enough systematic treatment in the literature but which typically can
have a substantial, if not detrimental, impact on the final
conclusions. One key component in data processing is to sort out
contradictions in the data, some obvious and some subtle.

A priori I did not expect this to be a part of the mystery that would
await me.
But that prior belief quickly shrank to $\varepsilon$ after reading the
first paragraph. The~authors started by emphasizing Pearson's (\citeyear{person20})
point that Fisher's likelihood is not useful for predicting future
observations or unobservables. Regardless of whether Fisher ever had
such an intention, this is an \textit{inference/prediction} issue. The~authors then immediately stated that existing efforts in generalizing
Fisher's likelihood inferences with unobservables run into the problem
of not having ``explicit forms'' due to the difficult in integration.
But that is squarely a \textit{computational/calculus} issue. Putting
aside the vast literature on the EM algorithm and related computational
methods that have successfully dealt with this very computational issue
in many common applications (see the overview by van Dyk and Meng
(\citeyear{VM09}) and other papers in the coming theme issue on EM in this
journal), I~am mystified by the logic and aims here---which issue do
the authors intend to address? Both?

Of course this could actually be a sign of a great mystery novel,
enticing the reader
from the very beginning, with multiple seemingly related or unrelated
lines to pursue, and a Holy Grail at the end---a gigantic~H! (Clearly
I am still in my Dan Brown mood, though hopefully this time the Holy
Grail is more than a legend.)

The~data processing indeed took much longer than I expected, mainly
because the ``unobservables'' that I need to infer, from a number of
mystic symbols whose meaning can only be surmised retrospectively to
reasons that can explain the authors' conviction that their
h-likelihood methods have been misunderstood by almost all the
discussants, since Lee and Nelder (\citeyear{LN1996}).

It is already 6:39 pm, Sunday. So let me get to the three main
storylines as I comprehend. The~first two lines are generally well
understood, so I shall reflect on them briefly. The~third line, which
is the most controversial, namely, h-likelihood inference for
unobservables, touches upon some fundamental issues about statistical
inference and prediction, and turns out to have at least one unexpected
intriguing property, at least to me. Therefore, the rest (three
quarters of the) discussion attempts to provide an explanation of this
controversy to a general audience, along with some ramifications and
thoughts it generates. Indeed, if a reader is in a rush to catch \textit
{Angles and Demons}, as my teenagers were, the reader should just skip
the following section, which contains no real enlightenment or
entertainment, other than some shameless self-advertisements and
academic quibbles.

\section{Two Uncontroversial Storylines}\label{sec3}
%
\subsection{Line One: Unobservables are Useful for~Modeling}

Much of the authors' Section~\ref{sec1} and Section~\ref{sec2} were devoted to arguing
and demonstrating the usefulness of unobservables for statistical
modeling. Other than the authors' preference for using \textit
{unobservables} as the all-encompassing term instead of the more common
term \textit{missing data} (though I agree that ``unobservables'' is
semantically more appropriate), the same message has been repeatedly
emphasized in the
literature, and it is indeed worthy of repeating. As I wrote in
``Missing Data: Dial M for ???'', a \textit{JASA} Y2K vignette (Meng,
\citeyear{meng2000}), ``The~topic of missing data is as old and as extensive as
statistics itself---after all, statistics is about knowing the
unknowns.'' Unable to outshine the summary there, I ask readers'
indulgence for a more extensive self-quotation. Below is the opening
paragraph of the same vignette, echoing well the authors' key emphases,
but with a more extended history (e.g., McKendrick's missing-data
modeling/formulation went back 1926; see Meng, \citeyear{meng1997}):

\begin{quote}
The~question mark is common notation for the missing data that
occur in most applied statistical analyses. Over the past century,
statisticians and other scientists not only have
invented numerous methods for handling
missing/incomplete data, but also have invented many forms of missing data,
including data augmentation, hidden states, latent variables,
potential outcome, and auxiliary variables.
Purposely constructing unobserved/unobservable variables
offers an extraordinarily flexible and powerful
framework for both scientific modeling and computation and
is one of the central statistical contributions
to natural, engineering, and social sciences. In parallel,
much research has been devoted to better understanding
and modeling of real-life missing-data mechanisms; that is,
the unintended data selection process that prevents us from
observing our intended data. This article is a very brief
and personal tour of these developments, and thus necessarily
has much missing history and citations. The~tour consists
of a number of Ms, starting with a historic story of the mysterious method
of McKendrick for analyzing an epidemic study and its link to the EM algorithm,
the most popular and powerful method of the twentieth century for
fitting models involving missing data and latent variables.
The~remaining Ms touch on theoretical, methodological and practical aspects
of missing-data problems, highlighted with some common applications
in social, computational, biological, medical and physical sciences.
\end{quote}

No further discussion seems necessary because this is a point on which
apparently most agree;
indeed, almost all the positive comments on Lee and Nelder (\citeyear{LN1996}) were
on praising
their promotion and formulation of models via unobservables.

\subsection{Line Two: H-likelihood for Fixed Parameter}
The~authors' Section~\ref{sec3} is where I saw the definition of h-likelihood
for the first time.
Using the authors' initial notation, $y$ denotes observed data, $\theta
$ is the fixed
parameter, and $v$ I infer is what the authors regarded as a random
``unobservable.'' The~h-loglikelihood is simply defined as $h(\theta, v)=\log f_\theta
(y, v)$ where
$f_\theta(y, v)$ is the joint probability distribution/\break density of $\{
y, v\}$.

In the rejoinder of Lee and Nelder (\citeyear{LN1996}), the authors argued that the
definition
of h-likelihood is as logical as Fisher's likelihood. I agree. In fact,
this point was
well recognized in Berger and Wolpert's (\citeyear{BW1988}) monograph on likelihood
principle (LP) where they wrote (page 21.2), ``\ldots the LP should be
formulated in such a way that $\theta$ consists of \textit{all} unknown
variables and parameters that are relevant to the statistical problem.'' (Emphasis is original.) They proceeded to devote an entire section to
the successes and challenges in extending the LP to include what they
call ``unobservable variables,'' just as in the authors' formulation. In
fact, in addition to the observable $X$, they wrote (pages~\mbox{36--37})
$\theta=(z; \omega)=(y,w;\xi,\eta)$, ``where $z=(y,w)$ is the value
of an unobservable variable $Z$ with $y$ being of interest and $w$
being a nuisance variable, and where $\omega=(\xi, \eta)$ is the
parameter that determines the distribution of both $X$ and $Z$, with
$\xi$ being of interest and $\eta$ being a nuisance parameter.'' This
quote shows that Berger and Wolpert's (\citeyear{BW1988}) definition is the same as
the authors', other than it takes a more explicit form by recognizing
two different kinds of unobservables, $y$~and~$w$, just as we often
distinguish between primary parameter $\xi$ and nuisance parameter~$\eta$.

The~key question here, therefore, is \textit{what to do with it} once it
is defined. I shall
discuss this point in Section~\ref{sec5}. Here it suffices to note that the
authors' initial proposal to maximize $h(\theta, v)$ jointly over $\{
\theta, v\}$, which they label MHLE (maximum h-likelihood estimation)
as in Section 2.2 of Lee and Nelder (\citeyear{LN1996}), can clearly lead to grossly
inconstant or even meaningless estimators if it is taken as a \textit
{general procedure}. This was pointed out by the majority of the
discussants of Lee and Nelder (\citeyear{LN1996}); as the authors stated later in
the rejoinder of Lee and Nelder (\citeyear{LN2004}), ``The~discussion was a disaster
because everybody took the worst possible case of binary data and
described difficulties with it. Nobody said it worked in other cases.'' The~example of Bayarri et al. (\citeyear{BDK1988}), reviewed in authors' Section~4.2,
demonstrated that the defect has little to do with binary data.

Indeed, earlier Little and Rubin (\citeyear{LR1983}) provided four examples, three
using standard univariate or bivariate (regression) normal models and
one with a censored exponential model, to show that MHLE (though of
course not in that term since Little and Rubin, \citeyear{LR1983} predates Lee and
Nelder, \citeyear{LN1996}) resulted in seriously flawed/inconsistent estimators,
unless the amount of missing data is (asymptotically) negligible. The~underlying issue is essentially the same as with the well-known
Neyman--Scott problem (Neyman and Scott, \citeyear{ns1948}). The~message here is
loud and clear: maximizing over unobservable/missing data, \textit{in
general}, is not a valid method.

Evidently, the message has been appreciated by the authors, as they now
make it explicit that for the ``fixed parameters,'' their method is the
same as Fisher's MLE, that is, maximizing the marginal log likelihood
$\ell(\theta)=\log f_\theta(y)$. This certainly should help to avoid
the type of mis-communications the authors described in the paper
(e.g., about Rubin and Little's \citeyear{LR2002} comments). But this also means
that no further discussion is needed either because there is no new
advance here.

However, for the sake of discussion, let me pick up on the authors'
statement that ``We view the marginal likelihood as an adjusted profile
likelihood eliminating nuisance unobservables $v$ from the
h-likelihood.'' The~issue is not much of the re-labeling itself, but
rather that by making such a statement, the authors might be in danger
of falling into the same trap that they have correctly warned others to
avoid. The~authors' ``adjusted profile h-likelihood (APHL),'' as far as
I am able to understand, simply uses a Laplace approximation to replace
the integration called for by Bayesian marginalization (for nuisance
parameter/unobservables). Whereas such an approximation indeed is very
useful and appealing for practical purposes \textit{when the
approximation is reasonable}, it does not constitute a \textit
{principled} statistical method in its own right unless a sound \textit
{inferential principle} is articulated for the approximation itself.
Without such a principle, its performance can only be judged by how
close the approximation is to the Bayesian target it approximates. In
this sense, comparisons such as those given in the authors' Figure~2
say little about the merit of the \mbox{h-likelihood} methods, but only
reconfirm the usefulness of the Laplace approximations, or demonstrate
the impact of the prior (which of course is not a part of the
h-likelihood formulation). In other words, mixing a \textit
{computation/approximation method} with a \textit{statistical method} is
as troublesome to me as mixing an \textit{estimation method} with a
\textit{statistical model} is to the authors (and to me of course).

Enough painless/itchless quibbles; let us get to the heart of the
authors' proposal, that is, making inference about the unobservables
via h-likelihood!

\section{What are the Principles Behind the H-likelihood Methods?}\label{sec4}

\subsection{Distinguishing Likelihood Principle, Likelihood
Inference, and MLE}
The~authors invoked several times the likelihood principle (LP) to
justify their h-likelihood methods. But all the LP says, broadly
speaking, is that if two data sets lead to the same likelihood, then
they contain the same information, assuming the underlying model for
each data set is correctly specified. The~LP eliminates any procedure
that violates it, but it says nothing about how to conduct a \textit
{likelihood inference}. As Berger and Wolpert [(\citeyear{BW1988}), Chapter 5] put
it, ``The~LP strikes us as correct, and behaving in violation of it
would be a source of considerable discomfort. Yet the LP does not tell
one what to do (although insisting on methods based on the observed
likelihood function certainly reduces the possibilities).''

Indeed, there is a long list of methods in the domain of ``utilization
of the likelihood function,'' too long even for Berger and Wolpert's
(\citeyear{BW1988}) monograph. I shall avoid repeating Berger and Wolpert's argument
that the full Bayesian inference is actually \textit{the most
principled} likelihood inference, since clearly the authors' intention
here is to achieve what Bayesian methods achieve but without adopting
the Bayesian philosophy; or, to self quote again (Meng, \citeyear{meng2008}),
``enjoying the Bayesian fruits without paying the B-club fee.'' But it
is worthwhile to re-emphasize that the notion of \textit{likelihood
inference} is a very elusive one---any method that does not violate LP
can be legitimately included (see Berger and Wolpert, \citeyear{BW1988}).

In contrast, maximal likelihood estimation (MLE) is a well-defined
method, telling us exactly what to do with the likelihood function. It
is this specific method that the authors' MHLE mimics. The~aforementioned counterexamples demonstrate clearly that in general this
imitation is only mathematical. The~key question then is whether it is
possible to find a set of useful and general conditions under which the
imitation is more fundamental, that is, under which MHLE preserves the
underlying properties of MLE that guarantee its validity and
efficiency? The~answer turns out to be an intriguing ``yes and NO.'' But
before we get to that punch line, we will need the wisdom of an old
friend, Mr. Bartlett.

\subsection{Do Bartlett Identities Hold for H-likelihood?}
Finding the \textit{most likely} parameter value that could have
produced the observed data
is intuitively very appealing---what else could be better? But of
course as statisticians we know such reasoning by itself is flawed,
because it puts us squarely in the hands of the Devil of Overfitting!
There is clearly much more to Fisher's MLE than this flawed intuition.

Probabilistically, a backbone of Fisher's ML\break method is the Bartlett
identities, especially the first two. That is, for the (marginal)
log-likelihood $\ell(\theta;y)$, under the usual regularity condition
that the support of $f_\theta(y)$ does not depend on $\theta\in
\Theta$, we have
%
\begin{equation}\label{eq:bart1}
E_\theta\biggl[\dfrac{\partial\ell(\theta;y)}{\partial\theta
}\biggr]=0\quad  \forall  \theta\in\Theta,\vspace*{-7pt}
\end{equation}
\begin{eqnarray}
\label{eq:bart2}
\hspace*{10pt}&&E_\theta\biggl[\dfrac{\partial^2 \ell(\theta;y)}{\partial\theta
^2}\biggr]+ E_\theta\biggl[\biggl(\dfrac{\partial\ell(\theta
;y)}{\partial\theta}\biggr)\biggl(\dfrac{\partial\ell(\theta
;y)}{\partial\theta}\biggr)^\top\biggr]\hspace*{-10pt}\nonumber\\[-8pt]\\[-8pt]
&&\quad =0\quad \forall  \theta
\in\Theta\nonumber
\end{eqnarray}
where $E_\theta$ denotes the expectation under $f_\theta(y)$. Here
identity (\ref{eq:bart1}) guarantees that the normal/score equation
underlying the MLE method,
%
\begin{equation}\label{eq:score}
S(\theta; y)\equiv\frac{\partial\ell(\theta;y)}{\partial\theta}=0,
\end{equation}
is an \textit{unbiased estimating equation}. Identity (\ref{eq:bart2})
is the basis for
the asymptotic efficiency of MLE (under regularity conditions, of
course) because it reduces the general ``sandwich'' variance formula to
the inverse of Fisher information, the Cram\'{e}r--Rao lower bound.

For these reasons, generalizations of (maximal) likelihood methods have
largely tried to
preserve these two identities, such as with the quasi-likelihood method
(e.g., McCullagh and Nelder, \citeyear{MN1989}, Chapter 9); see Mykland (\citeyear{M94}, \citeyear{M99}) for
other examples. It is therefore difficult to imagine that the issue of
preserving them has not been investigated in general in the context of
h-likelihood, given it is essentially a minimal requirement; indeed,
when Engle and Keen, the lead discussants of Lee and Nelder (\citeyear{LN1996}),
wrote, ``\ldots the usual first- and second moment properties exactly hold
for $h$-scores, for example, for normal-normal and Poisson-Gamma models\ldots'' I believe they were referring to the two identities above. I
therefore surmise that it is my haphazardly selective reading that
makes the existing investigations unobservable to me. So I must offer
my apologies to anyone, especially the authors, if I am reinventing the
wheel below. But in any case I hope the material presented in the rest
of this discussion will help to establish a firmer theoretical ground
for investigating the virtues and limitations of MHLE and other
h-likelihood methods.

Specifically, as we all know, identities (\ref{eq:bart1}) and (\ref
{eq:bart2}) are consequences of
%
\begin{eqnarray}\label{eq:cont}
\int_{\Omega_y} e^{\ell(\theta;y)}\mu(dy) = \int_{\Omega_y}
f_\theta(y) \mu(dy) = 1\nonumber\\[-8pt]\\[-8pt]
   \eqntext{\forall  \theta\in\Theta}
\end{eqnarray}
by repeatedly differentiating under the integral sign with respect to
$\theta$, which is legitimate when the support $\Omega_y$ is free of
$\theta$ (and assuming the usual continuous differentiability of $\ell
(\theta;y)$ as a function of $\theta$; such conditions will be
assumed below whenever needed). For log $h$-likelihood, $h(\theta
,v;y)=\log f_\theta(y,v)$, clearly we still have
\begin{eqnarray}\label{eq:hcont}
\quad &&\int_{\Omega_{y,v}} e^{h(\theta,v;y)}\mu(dy, dv)\nonumber\\[-8pt]\\[-8pt]
&&\quad  = \int_{\Omega
_{y,v}} f_\theta(y,v) \mu(dy, dv) = 1\quad   \forall  \theta\in
\Theta.\nonumber
\end{eqnarray}
However, whereas we can still take (partial) derivatives with respect
to $\theta$ on both sides of (\ref{eq:hcont}) to arrive at useful
identities, obviously taking the derivative of both sides with respect
to $v$ would produce $0=0$. This death of the old trick signifies a key
difference between the $h$-likelihood and Fisher's likelihood, even if
we put aside cases where $v$ is discrete and hence taking derivatives
is not even an option. Here we remark that unlike Fisher's likelihood
where discrete parameters are rare (other than with model selection
problems), discrete unobservable/missing data are common which poses an
additional challenge to the MHLE method. But clearly the authors'
current proposal focuses on continuous $v$, so we will proceed in this setting.

%

%
\section{Encouraging News: H-likelihood is~Bartlizable}\label{sec5}
%
\subsection{Necessary and Sufficient Conditions for Bartlett Identities}
Without the old trick, we have to directly investigate if and when
(\ref{eq:bart1}) and (\ref{eq:bart2}) can be extended to $h$-likelihood.
Specifically, when we let $\phi=\{\theta, v\}$, and write
%
\begin{equation}\label{eq:iden}
h(\phi;y)=\log f_\theta(y|v)+\log f_\theta(v),
\end{equation}
we see the ``troublemaker'' is the second term
because for the first term, $v$ plays the same role of a fixed
parameter for the conditional distribution $f_\theta(y|v)$, and hence
the old trick of differentiating
under integration is applicable. In particular, as an application of
(\ref{eq:bart1}) and (\ref{eq:bart2}) when conditioning on $v$ and
assuming the support of $f_\theta(y|v)$ does not depend on either
$\theta$ or $v$, we have, for any $\theta\in\Theta$,
%
\begin{equation}\label{eq:bartv1}
E_\theta\biggl[\frac{\partial\log f_\theta(y|v)}{\partial\phi
}\Big|  v \biggr]=0,\vspace*{-7pt}
\end{equation}
%
\begin{eqnarray}
\label{eq:bartv2}
\hspace*{16pt}&&E_\theta\biggl[\frac{\partial^2\log f_\theta(y|v)}{\partial\phi
^2}\Big|  v \biggr]\hspace*{-16pt}\nonumber\\
&&\qquad {}+
E_\theta\biggl[\biggl(\frac{\partial\log f_\theta(y|v)}{\partial
\phi}\biggr)\biggl(\frac{\partial\log f_\theta(y|v)}{\partial\phi
}\biggr)^\top\Big|  v \biggr]\\
&&\quad =0.\nonumber
\end{eqnarray}
Consequently, under the additional assumption that the support of
$f_\theta(v)$ does not depend on $\theta$, (\ref{eq:iden}) and (\ref
{eq:bartv1}) imply that, for any $\theta\in\Theta$,
%
\begin{eqnarray}\label{eq:barth1}
E_\theta\biggl[\frac{\partial h(\phi;y)}{\partial\phi}
\biggr]&=&E_\theta\biggl[\frac{\partial\log f_\theta(v)}{\partial\phi
}\biggr]\nonumber\\
 &=&\pmatrix{
0 \vspace*{2pt}\cr
E_\theta\biggl[\dfrac{\partial\log f_\theta(v)}{\partial v}\biggr]}\\
&\equiv&
\pmatrix{
0 \vspace*{2pt}\cr
\displaystyle\int_{\Omega_v}\frac{\partial f_\theta(v)}{\partial v}\mu(dv)}.\nonumber
\end{eqnarray}
Furthermore, noting that the cross terms in the\break quadratic expansion
below are zero by first conditioning on $v$, we have from (\ref
{eq:iden})--(\ref{eq:bartv2}),
\begin{eqnarray}
\label{eq:barth2}
\hspace*{10pt}&&E_\theta\biggl[\frac{\partial^2 h(\phi;y)}{\partial\phi^2}\biggr]+
E_\theta\biggl[\biggl(\frac{\partial h(\phi;y)}{\partial\phi
}\biggr)\biggl(\frac{\partial h(\phi;y)}{\partial\phi}
\biggr)^\top\biggr] \nonumber\\
&&\quad = E_\theta\biggl[\frac{\partial^2 \log f_\theta(v)}{\partial\phi
^2}\biggr]\nonumber\\[-8pt]\\[-8pt]
&&\qquad {}+
E_\theta\biggl[\biggl(\frac{\partial\log f_\theta(v)}{\partial\phi
}\biggr)\biggl(\frac{\partial\log f_\theta(v)}{\partial\phi
}\biggr)^\top\biggr]\nonumber\\
&&\quad \equiv\pmatrix{
A & B\vspace*{2pt}\cr
B^{\top}& C }.\nonumber
\end{eqnarray}
In the above expression,
%
\begin{eqnarray}\label{eq:aa}
A&=& E_\theta\biggl[\frac{\partial^2 \log f_\theta(v)}{\partial
\theta^2}\biggr]\nonumber\\[-8pt]\\[-8pt]
\hspace*{3pt}&&{}+
E_\theta\biggl[\biggl(\frac{\partial\log f_\theta(v)}{\partial
\theta}\biggr)\biggl(\frac{\partial\log f_\theta(v)}{\partial
\theta}\biggr)^\top\biggr]=0\hspace*{-19pt}\nonumber
\end{eqnarray}
by applying (\ref{eq:bart2}) to $\log f_\theta(v)$. For the term $B$,
one can easily verify that
%
\begin{eqnarray}
\label{eq:bb}
B&=& E_\theta\biggl[\frac{\partial^2 \log f_\theta(v)}{\partial
\theta\,\partial v}\biggr]\nonumber\\
&&{}+
E_\theta\biggl[\biggl(\frac{\partial\log f_\theta(v)}{\partial
\theta}\biggr)\biggl(\frac{\partial\log f_\theta(v)}{\partial
v}\biggr)^\top\biggr]\\
&=&\frac{\partial}{\partial\theta}\biggl\{ E_\theta\biggl[
\biggl(\frac{\partial\log f_\theta(v)}{\partial v}\biggr)^{\top}
\biggr]\biggr\}\quad  \forall  \theta\in\Theta,\nonumber
\end{eqnarray}
and hence it will also be zero if $E_\theta[\frac
{\partial\log f_\theta(v)}{\partial v}]=0$ for all $\theta\in
\Theta$. Finally, simple algebra shows
%
\begin{eqnarray}\label{eq:cc}
C&=& E_\theta\biggl[\frac{\partial^2 \log f_\theta(v)}{\partial
v^2}\nonumber\\
&&\hspace*{18pt}{}+\biggl(\frac{\partial\log f_\theta(v)}{\partial v}\biggr)
\biggl(\frac{\partial\log f_\theta(v)}{\partial v}\biggr)^\top
\biggr]\\
&\equiv&
\int_{\Omega_v}\frac{\partial^2 f_\theta(v)}{\partial v^2}\mu(dv).\nonumber
\end{eqnarray}

Combining (\ref{eq:barth1})--(\ref{eq:cc}) yields the following
straightforward but key result.

\begin{theorem}\label{teo1}
Let $h(\phi;y)=\log f_\theta(y,v)$ be a log
h-likelihood where $\phi=\{\theta,v\}$, $\theta\in\Theta$ is the
model parameter, $v$ is a continuous unobservable with density
$f_\theta(v)$ with respect to a measure $\mu$, and let $\mathcal
{S}_\theta
(v)= \frac{\partial\log f_\theta(v)}{\partial v}$. Furthermore,
assume the support of $f_\theta(y|v)$ does not depend on either
$\theta$ or $v$ (almost surely with respect to $\mu$), the support of
$f_\theta(v)$, denoted by $\Omega_v$, is free of $\theta$, and all
continuity and differentiability conditions hold whenever needed. Then
the first Bartlett identity holds for the h-likelihood, that is
%
\begin{equation}\label{eq:h1}
E_\theta\biggl[\frac{\partial h(\phi;y)}{\partial\phi}\biggr] =0
 \quad \forall  \theta\in\Theta
\end{equation}
if and only if
%
\begin{equation}\label{eq:hiden1}
\hspace*{20pt}E_\theta[\mathcal{S}_\theta(v)]\equiv\int_{\Omega_v}\frac
{\partial
f_\theta(v)}{\partial v}\mu(dv)=0\quad   \forall \theta\in
\Theta.\hspace*{-20pt}
\end{equation}
Assuming (\ref{eq:hiden1}), then the second Bartlett identity holds
for the h-likelihood; that is,
%
\begin{eqnarray}\label{eq:h2}
&&E_\theta\biggl[\frac{\partial^2 h(\phi;y)}{\partial\phi^2}\biggr]\nonumber\\
&&\qquad {}+
E_\theta\biggl[\biggl(\frac{\partial h(\phi;y)}{\partial\phi
}\biggr)\biggl(\frac{\partial h(\phi;y)}{\partial\phi}
\biggr)^\top\biggr]\\
&&\quad =0\quad  \forall  \theta\in\Theta\nonumber
\end{eqnarray}
if and only if
%
\begin{eqnarray}\label{eq:hiden2}
&&E_\theta\biggl[\frac{\partial\mathcal{S}_\theta(v)}{\partial v}+
\mathcal{S}_\theta
(v) \mathcal{S}_\theta^\top(v)\biggr]\nonumber\\[-8pt]\\[-8pt]
&&\quad \equiv\int_{\Omega_v}\frac
{\partial
^2 f_\theta(v)}{\partial v^2}\mu(d v)=0\quad   \forall  \theta\in
\Theta.\nonumber
\end{eqnarray}
\end{theorem}

%
\subsection{Yes: It is Easy for H-likelihood to Produce
``Un-sandwiched'' Estimating Equation}\label{sec5.2}

Theorem~\ref{teo1} is somewhat remarkable because the necessary and sufficient
conditions (\ref{eq:hiden1}) and (\ref{eq:hiden2}) are determined
purely by the marginal distribution of the unobservable $v$, and hence
they are easy to check. For example, in Bayarri's example quoted by the
authors, the marginal density of the unobservable $u$ is exponential
with mean $\lambda=\theta^{-1}$. Consequently, $\mathcal{S}_\theta
(u)=-\theta$, and hence condition (\ref{eq:hiden1}) is violated for
all $\theta>0$, as is condition (\ref{eq:hiden2}). This means that
whenever $u$ is used for the h-likelihood, the resulting h-score will
never form an unbiased estimating equation, regardless of the model for
$f_\theta(y|u)$! Indeed, we have seen from the authors' Section~4.2
that the corresponding MHLE leads to meaningless estimates.

In contrast, when we use $v=\log u$, $f_\theta(v)=\theta e^{v-\theta
e^v}$, and hence $\mathcal{S}_\theta(v)=1-\theta e^{v}=1-u/\lambda$
and $\mathcal{S}
_\theta^{\prime}(v)+\mathcal{S}_\theta^2(v)=
-\theta e^v + (1-\theta u)^2 = -u/\lambda+ (u-\lambda)^2/\lambda^2$.
Both conditions (\ref{eq:hiden1})
and (\ref{eq:hiden2}) then follow trivially because $E_\theta
(u)=\lambda$ and $V_\theta(u)=\lambda^2$.
Consequently, the authors' h-score is not only an unbiased estimating
equation but also an ``optimal'' one in the sense that we do not need
the usual ``sandwich'' formula, but only the Hessian matrix, for ``valid'' variance estimation. Unfortunately, I have to put both ``optimal'' and
``valid'' in quotes because of the \textit{bad news} I will deliver in
the next section. But as far as for preserving Bartlett identities
goes, \textit{which by itself does not guarantee valid statistical
inferences}, I can share the authors' optimism for the future of MHLE,
especially because of the following somewhat even more surprising
result, which says that conditions (\ref{eq:hiden1}) and (\ref
{eq:hiden2}) hold quite easily for many unobservables or their simple
transformations.

\begin{theorem}\label{teo2}
Under the same setting as in Theorem~\ref{teo1}, suppose the support of
$f_\theta(v)$, $\Omega_v \subset R^d$, takes a rectangle form,
$\Omega_v=\prod_{j=1}^d [a_j, b_j]$, where $a_j$ or $b_j$ is
permitted to take the value of $+\infty$ or $-\infty$. Let $\partial
\Omega_v$ be the boundary set of $\Omega_v$ (i.e., the set of all
points whose coordinates contain at least one $a_j$ or $b_j$), and
assume the dominating measure $\mu$ is the Lebesgue measure on~$R^d$.
We then have:
\begin{longlist}[(II)]
\item[(I)] If $f_\theta(v)=0$ for all $v\in\partial\Omega_v$, then
condition (\ref{eq:hiden1}) holds, and hence the first Bartlett
identity (\ref{eq:h1}) holds.

\item[(II)] If in addition $\frac{\partial f_\theta(v)}{\partial
v}=0$ also holds for all \mbox{$v\in\partial\Omega_v$}, then condition
(\ref{eq:hiden2}) holds, and hence the second Bartlett identity (\ref
{eq:h2}) holds.
\end{longlist}
\end{theorem}

\begin{pf} For (I), because of (\ref{eq:hiden1}), if $v$ is
univariate, that is, if $d=1$, then
%
\begin{eqnarray}\label{eq:h1p}
\int_{\Omega_v}\frac{\partial f_\theta(v)}{\partial v}\,d v
&=&\int_{a_1}^{b_1}d f_\theta(v)\nonumber\\
&=&f_\theta(b_1)-f_\theta(a_1)\\
&=&0,\nonumber
\end{eqnarray}
under our assumption that $f_\theta(v)$ vanishes on the boundary. For $d>1$,
we apply the same argument to each of the $d$ integrations that form
the leftmost vector in (\ref{eq:h1p}), that is, $\int_{\Omega
_v}\frac{\partial f_\theta(v)}{\partial v_k}\,d v,  k=1,\ldots, d$,
by integrating with respect to $v_k$
first to conclude that it is zero for all $\theta$.

For (II), we first note that for any $\{k, s\}$,
%
\begin{eqnarray}\label{eq:hp3}
I_{k,s}&\equiv&\int_{\Omega_v}\frac{\partial^2 f_\theta
(v)}{\partial v_k\,\partial v_s}\,d v\nonumber\\[-8pt]\\[-8pt]
&=&\int_{\Omega_v}\frac{\partial
}{\partial v_k}\biggl(\frac{\partial f_\theta(v)}{\partial v_s}
\biggr)\,d v.\nonumber
\end{eqnarray}
Hence, using the same argument as above but with $f_\theta(v)$
replaced by $\frac{\partial f_\theta(v)}{\partial v_s}$, we can conclude
$I_{k,s}=0$ for all $k, s=1,\ldots, d$. Consequently, condition
(\ref{eq:hiden2}) holds.
\end{pf}

What this result says is that as long as the marginal density of the
unobservable $v$ vanishes on the boundary of its support, the first
Bartlett identity holds for h-likelihood. In addition, if its
derivative also vanishes on the boundary, then the second Bartlett
identity holds. This provides an even easier way to verify Bayarri's
example. For the original unobservable~$u$, $f_\theta(u)=\theta
e^{-\theta u}$, with boundary points $u=0$ and \mbox{$u=\infty$}. But since
$f_\theta(0)=\theta$, the vanishing condition is violated as long as
$\theta>0$. In contrast, for $v=\log u$,
$f_\theta(v)=\theta e^{v-\theta e^v}$, with boundary points $v=-\infty
$ and $v=+\infty$.
It is easy to see that $f_\theta(-\infty)=f_\theta(+\infty)=0$ for
all $\theta$. Furthermore, since
\[
\frac{\partial f_\theta(v)}{\partial v}=\theta(e^{v-\theta
e^v}-\theta e^{2v-\theta e^v}),
\]
the derivative also vanishes at both $v=-\infty$ and $v=\infty$.
Therefore, both Bartlett identities hold for \mbox{h-likelihood} when $v=\log
u$ is used as the unobservable.
For simplicity, we will label the process of finding a transformation
that makes Bartlett identities hold \textit{Bartlization}
(``Bartlettlization'' is too much of a tongue twister!).

An astute reader may have noticed that I did not say that failing the
vanishing condition is the reason for the failing of the Bartlett
identities for the original scale~$u$. The~vanishing condition is
sufficient, but not necessary. This can easily been seen in (\ref
{eq:h1p}), which only requires $f_\theta(a_1)=f_\theta(b_1)$. Indeed,
the Bartlett identity fails for the original scale $u$ precisely
because $f_\theta(u=+\infty)=0$ but $f_\theta(u=0)=\theta$, and
hence $E_\theta[\mathcal{S}_\theta(u)]= 0-\theta=-\theta$, as verified
directly previously. A necessary and sufficient condition via
integration on $\partial\Omega_v$ is not hard to obtain, but it
requires a bit more mathematical treatment than is needed for most
practical applications, for which Theorem~\ref{teo2} is adequate. Here we just
mention that we can generalize Theorem~\ref{teo2} by allowing $\Omega_v$ to be
an arbitrary \textit{simply connected} manifold in $R^d$ (i.e., a
manifold with ``no hole''), and then invoke the generalized Stokes'
theorem (see Marsden and Tromba, \citeyear{MT2003}) to equate the integration of
$dw$ on $\Omega_v$ to that of $w$ on the boundary $\partial\Omega_v$
where $w$ is a so-called $d-1$ differential form which can be taken
in terms of $f_\theta(v)$ or its derivative as needed.

The~authors stated in the rejoinder of Lee and Nelder (\citeyear{LN2004}) that ``We
do not say that the current h-likelihood method will always perform the
best, but we believe that it can always be modified to give an
improvement, as has been done with Fisher's likelihood method.'' I~believe the alluded-to improvements lie in using higher order Bartlett
identities, such as the third identity for ``Bartlett correction'' for
the likelihood ratio tests (e.g., McCullagh, \citeyear{MC1987}). Clearly Theorem \ref{teo1}
and Theorem \ref{teo2} have their higher order generalizations, but it is
already 9:14 pm of the \textit{second} Sunday. My teenagers' visit to Dr.
Langdon is already postponed for another week, so I had better leave
such generalizations to a future paper. More importantly, as much as I
am enjoying discovering the ``Bartlizability'' of \mbox{h-likelihood}, I do not
see a way to correct the more fundamental problem described in the next
section, which potentially makes ``Bartlett-corrected h-likelihood'' an
exercise that is literally just a homework exercise.

\section{Bad News and A Puzzle: Fishy~or~Fiducial?}\label{sec6}
%
\subsection{NO: It is Hard for log H-likelihood to be
Summarizable Quadratically}

Having the Bartlett identities is only a part of the story. What it
guarantees is that \textit{if the log h-likelihood can be approximated
quadratically}, then the mode and the Hessian matrix derived from it
will provide an approximately correct estimator and its associated
(inverse) variance. To examine this issue more clearly, let us mimic
the formal asymptotic argument behind the estimating equation approach
which relies on the
expression
%
\begin{equation}\label{eq:taylor}
\hat\phi- \phi= I_h^{-1}(\theta)\mathcal{S}(\phi; y) + R,
\end{equation}
where $\hat\phi$ is the MHLE, $\mathcal{S}(\phi; y)=\frac{\partial
h(\phi
;y)}{\partial\phi}$ is the h-score, and $I_h(\theta)$ is the
h-likelihood extension of the expected Fisher information, the expected Hessian,
%
\begin{equation}\label{eq:fisher}
I_h(\theta)\equiv E_\theta\biggl[-\frac{\partial^2 h(\phi
;y)}{\partial\phi^2}\biggr].
\end{equation}
We emphasize here that unlike the original Fisher information,
$I_h(\theta)$ is not
generally guaranteed to be positive definite (so $I^{-1}_h(\theta)$
may not even exist) \textit{unless condition \textup{(\ref{eq:hiden2})} holds};
see Section~\ref{sec7} for an example.

Expression (\ref{eq:taylor}) by itself is tautological, because there
is always an $R$ to make it hold; in particular it can be derived from
a remainder term in the Taylor expansion of $\mathcal{S}(\hat\phi
;y)-\mathcal{S}(\phi
;y)$. However, \textit{when $R$ is (asymptotically) negligible}, (\ref
{eq:taylor}) allows us to conclude that the distribution of $\hat\phi
- \phi$ can be approximated by that of $T(\theta;y)\equiv
I_h^{-1}(\theta)\mathcal{S}(\phi; y)$ which has mean zero when the first
Bartlett identity holds
and variance $I^{-1}_h(\theta)$ when the second Bartlett identity holds.

When $h$ is a regular Fisher's likelihood, under regularity conditions,
the $R$ term is asymptotically negligible compared with the first term
on the right-hand side of (\ref{eq:taylor}). A key reason for this is
the \textit{accumulation of information} as we collect more data;
eventually we will have zero uncertainty about the parameter, at least
in theory. Unfortunately, for h-likelihood, this cannot be true \textit
{in general} even in theory because no matter how much data we
accumulate, it cannot possibly eliminate the uncertainty, say, in
predicting a \textit{future outcome}, such as in the authors' Example~4.
This lack of accumulation of information for unobservables is
essentially the key problem pointed out by multiple discussants (e.g.,
both lead discussants) of Lee and Nelder (\citeyear{LN1996}), with both theoretical
and empirical examples.

Without the accumulation of information to justify the central limit
theorem or the law of large numbers, we actually will run into two
problems with the standard asymptotic arguments
for (\ref{eq:taylor}), even if the first two Bartlett identities hold.
The~most obvious and critical one is that since $R$ is not negligible,
we cannot approximate the distribution of $\hat\phi- \phi$ by that
of $T(\theta;y)=I_h^{-1}(\theta)\mathcal{S}(\phi; y)$; indeed,
without $R$
being negligible, the MHLEs are not guaranteed to be consistent, as in
all examples of Little and Rubin (\citeyear{LR1983}). It is of critical importance
to stress that the Bartlizable property of h-likelihood itself has
little bearing on the issue of being quadratically summarizable, that
is, the $R$ term being negligible. Indeed, in all normal examples of
Little and Rubin (\citeyear{LR1983}),
the h-likelihood is naturally Bartlized because clearly the normal
density and any of its derivatives vanish on the boundary of its
support, yet the MHLE produces inconsistent estimators because of the
nonnegligibility of the $R$ term. The~more subtle one is that
regardless of whether $R$ is negligible or not, we may not be able to
justify the usual normal approximation $T(\theta;y) \sim N(0,
I_h^{-1}(\theta))$, even if $T(\theta;y)$ has mean zero and variance
matrix~$I_h^{-1}(\theta)$. (Of course, when $R$ is not negligible, the
properties of $T$ are not really relevant.) Section~\ref{sec7} will illustrate
all these points via a simple but very informative example.

\subsection{And a Puzzle: The~Meaning of the H-distribution}
Even if $R$ is exactly zero and all Bartlett identities hold, the
h-likelihood method, as a method for predicting the unobservables $v$,
still faces a fundamental challenge. That is, what is the meaning of
the resulting distribution $f(v|y)$, which I shall term the \textit
{$h$-distribution} for obvious reasons? If one is willing to assume a
constant prior on $\theta$, then of course this has a Bayesian
interpretation as a posterior predictive distribution or an
approximation to it. But the authors specifically emphasized that they
did not want to specify a prior on $\theta$, for their goal is to
provide an alternative method to the Bayesian approach.

Some Bayesians may be agitated by having a\break method that is
mathematically or numerically equivalent, in general, to a Bayes method
(perhaps under a particular prior), but is labeled as something else. I
am much less troubled, provided that (1) the connection is clearly
spelled out, and (2) there is a well-articulated non-Bayesian principle
justifying the method. The~authors clearly have done (1), but for (2)
all I can find is authors' desire to conduct a \textit{probabilistic
inference} for $v$ without having to specify a prior for $\theta$. At
the conceptual level, I have the very same desire because of my
frustration, which I am sure some share, with the apparent
impossibility of constructing a truly ``noninformative'' prior (for
continuous parameters, at least). I also very much appreciate the
authors' emphasis that the ``plug-in'' empirical Bayes is not a
satisfying method, precisely because ``plug-in'' is an ad hoc method. So
indeed I was quite excited when I thought that the authors had found a
way to meaningfully specify a \textit{probabilistic} $f(v|y)$, without
considering $\theta$ as a random variable.

At a practical level, the authors did provide a number of
``well-specified'' h-distributions, either via (the implied) normal
approximation with mean and variance obtained from the MHLE/Hessian
matrix for $v$ or the APHL approximation by profiling out $\theta$.
But without spelling out the probabilistic meaning of such resulting
distributions, it is essentially impossible to answer the criticism
that the label of h-likelihood is a red herring because they are just
approximations to Bayesian solutions instead of the products of a
genuine competing method as claimed. More importantly, without knowing
what ``gold standard'' they aim to approximate, we have no meaningful
ways to evaluate how good the approximations are, or even to specify a
probabilistic evaluation mechanism; in what real or thought experiment
can it be realized?

Indeed, the lack of a distinct and justifiable meaning of the
h-distribution apparently has put the authors in an awkward position in
terms of demonstrating the merit of their methods. From the papers I
read, it appears that the authors have two kinds of comparisons. The~first is to compare an h-distribution to a Bayesian one, and to
``validate'' the h-distribution by showing how close it is to the
Bayesian counterpart. But this only strengthens the aforementioned
``red herring'' criticism, and provides evidence for---not against---the
kind of statements made by my twin brother quoted previously. Clearly
this is contrary to the authors' intention, and I believe is part of
the reasons for the continuing discrepancy between the authors'
enthusiasm for and others' reluctance toward the h-likelihood methods.

The~second type is something that I have not seen before, at least not
in academic publications. The~authors seem to take their methods as the
standard, and compared everything else to it, as suggested by the
statement, ``In the salamander data, among other methods considered,
the MCEM of\break Vaida and Meng (\citeyear{VaMe2005}) gives the closest estimates to the
h-likelihood estimators.'' Such comparisons would be meaningful if the
superiority of the h-likelihood results had already been demonstrated
either by theoretical proof (e.g., optimality of some sort) or by a
distinctive principle that is not subsumed or invalidated by accepted
ones. But even in such cases, the value of this type of comparison is
to demonstrate the performance of \textit{other methods}, not the merit
of the h-likelihood method itself.

%
\subsection{Fiducial Argument via Predictive Pivotal~Quantity?}
As I tried in vain to form a thought experiment that would meaningfully
define the h-distribution $f(v|y)$ without slipping into the Bayesian
mode, I looked hard into the authors' writings for clues about what
they had in mind. The~first clue came from Section 3.1 of Lee and
Nelder (\citeyear{LN1996}), where they showed that, in the context of the models
they were investigating, a log \mbox{h-likelihood} expression in their (3.2)
can be expanded into their expression (3.3) which is a quadratic term
$-(\tilde v - v)'D^*(\tilde v - v)/2$ plus a term that depends on $y$
only (their $\tilde v$ is the same as the $\hat v$ notation here). They
then wrote, ``\textit{Ignoring the constant term, which depends only on
$y$ and not on $v$, expressions \textup{(3.2)} and~\textup{(3.3)} imply that
\[
v|y \sim N(\tilde v, D^{* -1})
\]
would be a good approximation for the distribution of~$v|y$}.'' With apologies to the authors in case I misunderstand their notation or
there was a misprinting, this reasoning smells either fishy or
fiducial, depending on the meaning of ``the distribution of $v|y$.''

First, if by ``the distribution of $v|y$'' is meant the sampling
distribution of $v$ \textit{given both} $y$ and $\theta$, then the
reasoning underlying the above statement would contain the elementary
flaw of confusing a \textit{marginal} distribution of $X_1-X_2$ with the
\textit{conditional} distribution of $X_1-X_2$ given $X_1$. This is
because, even if the normal approximation is justified, the quadratic
term above is for the \textit{marginal} distribution of $\tilde v -v$,
as $v$ and~$\tilde v$, which is a function of $y$ only, vary \textit
{jointly} according to $f(y; v|\theta)$. [I switch the notation from
$f_\theta(y; v)$ to $f(y; v|\theta)$
to emphasize the conditioning on $\theta$, even though the latter
notation may imply that $\theta$ is a \textit{variable} being
conditioned upon, something the authors' approach aims to avoid.] This
marginal distribution clearly is not the same, in general, as the \textit
{conditional} distribution $f(\tilde v -v| \tilde v, \theta)$ or
$f(\tilde v -v|y, \theta)$ (note in general that these two
distributions are also different unless $y$ and $v$ are independent
given $\theta$). This can be most clearly seen from (\ref{eq:taylor})
where all the distributional calculations are with respect to the joint
distribution $f(y; v|\theta)$, \textit{not} the conditional
distribution $f(v|y;\theta)$.

Of course, this is unlikely to be what the authors intended, since
their goal is to capture $v|y$ without conditioning
on $\theta$. But the notation $f(v|y)$ has no definition or meaning
under the authors' joint modeling specification $f(y, v|\theta)$
because $\theta$ is treated as fixed. This brings me to the second
``smell,'' that is, the authors were invoking a fiducial-like argument,
by implicitly \textit{defining} their \textit{conditional h-distribution}
$v|y$ as the \textit{sampling marginal distribution} of $\hat v - v$
under the joint distribution $f(y,v|\theta)$, and getting rid of its
dependence on $\theta$ when $\hat v - v$ is (asymptotically) a \textit
{predictive pivotal quantity}, meaning that its distribution is free of
any unknowns.
We can also think of this way of eliminating the nuisance parameter
$\theta$ for the purpose of prediction as seeking \textit{predictive
ancillarity}, that is, a function of both $y$ and $v$ whose
distribution is free of~$\theta$. See the example in Section~\ref{sec7} for an
illustration.

%
\subsection{A Duality or Prestidigitation?}
The~second piece of evidence from the authors' writing seems to confirm
this interpretation.
In the comparisons of their methods with the Empirical Bayesian method,
they compared the Bayesian posterior predictive variance of $v|y$ with
the estimator obtained from the Hessian matrix. To make this comparison
more explicit, let us denote $\tau(\theta;y)= V(v|\theta;y)$ and
$e(\theta;y)=\hat v(y) - E(v|\theta;y)$. Then by the law of iterated
expectations (or the so-called EVE formula) and noting that $\hat v$ is
determined by $y$, we have
%
\begin{eqnarray}\label{eq:post}
V(v|y)&=&V(\hat v -v|y)\nonumber\\[-8pt]\\[-8pt]
&=& E[\tau(\theta;y)|y]+ V[e(\theta;y)|y],\nonumber\\
\label{eq:hvar}
\qquad  V(\hat v - v|\theta)&=&E[\tau(\theta;y)|\theta]+ V[e(\theta
;y)|\theta].
\end{eqnarray}
The~authors' argument seems to implicitly rely on a ``duality,'' that
is, the two mean terms on the right-hand sides of (\ref{eq:post}) and
(\ref{eq:hvar}) are (asymptotically or approximately) the same; so are
the two variance terms. That is, we can switch the required mean and
variance calculations under $f(y|\theta)$ in (\ref{eq:post}) to that
under $f(\theta|y)$ in (\ref{eq:hvar}).
Fisher's fiducial argument, as far as I can understand, aimed to
establish the validity of this switching on its own without viewing it
as an approximation to the Bayesian method (with a constant prior).
There is nothing wrong with invoking the fiducial argument (well,
actually there is but it depends on who one asks); indeed there has
been a recent surge of interest in it, especially in connection with
the ``generalized confidence'' approach [e.g., Hannig, Iyer and Patterson (\citeyear{HIP06}) and
Hannig (\citeyear{Han09})]. Perhaps the authors' approach is the next step, that
is, using the fiducial approach for prediction, not just for
estimation. But without being told explicitly about this switching, a
reader's reaction would be anybody's guess. A suspicion of
prestidigitation? A deja vu feeling of reading \textit{Deception Point}
instead of \textit{De Vinci Code}? Or even worse, an accusation of the
prosecutor's fallacy?

Finally, even if we buy the fiducial argument, it does not follow that
the left-hand side of (\ref{eq:hvar}) can be well approximated by (an
appropriate element of) the inverse of the Hessian matrix because of
the non-negligibility of the $R$ term, as discussed before. The~authors, of course, well recognized this, and hence invoked the APHL
method to approximate (define?) the h-distribution $f(v|y)$ instead of
relying on the normal approximation. While this approach indeed ``works
well,'' in the authors' example and in the example I am about to
present, I have to put ``works well'' in quotes when the success is
judged by comparing how close the h-distribution is to the posterior
predictive distribution under the constant prior. But I'd be happy to
remove the quotation marks if the evaluation is based on the
aforementioned pivotal predictive framework, because that is a
distinctive principle, regardless of whether one subscribes to it or not.

%

\section{Show and Tell: Estimation and Prediction with
Exponential Distribution}\label{sec7}

To illustrate various general points made in Sections~\ref{sec4}--\ref{sec6}, let us
consider a simple case where the data are an i.i.d. sample $y=\{
y_1,\ldots, y_n\}$ from an exponential distribution with mean $\lambda
$ with the unobservable being $u=y_{n+1}$, a future observation.
This example is different from Bayarri's two-level exponential model
because here we only have one level, as in the authors' Example 4. It
is hard to have faith in a method for multi-level hierarchical models
if it cannot handle single-level models.

\subsection{Why does the Original Scale Fail?}
As we discussed in Section~\ref{sec5.2}, when the exponential variable
$u=y_{n+1}$ is used as the unobservable, the Bartlett identities fail.
In the current setting, this can be seen directly by noting that (where
$\bar y_n$ denotes the sample mean of $\{y_1,\ldots, y_n\}$)
%
\begin{equation}\label{eq:expo1}
h(\lambda, u; y)= -(n+1)\log\lambda- \frac{n\bar y_n+u}{\lambda},
\end{equation}
which clearly does not have an internal mode because it is linear in
$u\ge0$. Indeed,
the h-score equation,
%
\begin{eqnarray}\label{eq:expo2}
S(\phi; y)&\equiv&\pmatrix{
\dfrac{\partial h}{\partial\lambda} \vspace*{2pt}\cr
\dfrac{\partial h}{\partial u} }\nonumber\\[-8pt]\\[-8pt]
&=&
\pmatrix{
-\dfrac{n+1}{\lambda}
+\dfrac{n \bar y_n +u}{\lambda^2}\vspace*{2pt}
\cr
-\dfrac{1}{\lambda}}= \pmatrix{
0\vspace*{2pt}\cr
0 },\nonumber
\end{eqnarray}
leads to the meaningless estimator $\hat\lambda=+\infty$.
Incidently, this is also an example that $I_h(\theta)$, as defined in
(\ref{eq:fisher}), is not nonnegative definite because the second
Bartlett identity fails. Specifically, by further differentiating the
expressions in (\ref{eq:expo2}), it is easy to verify that
\begin{eqnarray*}
I_h(\theta)&=& E\left[-\pmatrix{
\dfrac{n+1}{\lambda^2}-2\dfrac{n \bar y_n +u}{\lambda^3} & \dfrac
{1}{\lambda^2}\vspace*{2pt}\cr
\dfrac{1}{\lambda^2} & 0}\right]\\
&=&
\pmatrix{
\dfrac{n+1}{\lambda^2}& -\dfrac{1}{\lambda^2}\vspace*{2pt} \cr
-\dfrac{1}{\lambda^2} & 0}
\end{eqnarray*}
which clearly fails to be nonnegative definite.

\subsection{A Simple Transformation is All it Takes}
However, when the h-likelihood uses $v=\log(u)$ as unobservable,
it satisfies both conditions of Theorem~\ref{teo2} as verified in Section~\ref{sec5.2},
so the corresponding \mbox{h-likelihood} is Bartlized. To see this directly, because
%
\begin{equation}\label{eq:expov1}
\qquad h(\lambda, v; y)= -(n+1)\log\lambda- \frac{n\bar y_n+e^v}{\lambda}+v,
\end{equation}
the h-score equation becomes
%
\begin{eqnarray}\label{eq:expov22}
\frac{\partial h}{\partial\lambda}& =& -\frac{n+1}{\lambda}
+\frac{n \bar y_n +e^v}{\lambda^2}= 0,\nonumber\\[-8pt]\\[-8pt]
\frac{\partial h}{\partial v}& =& -\frac{e^v}{\lambda}+1
= 0.\nonumber
\end{eqnarray}
This delivers the correct MLE for $\lambda$, $\hat\lambda= \bar
y_n$, and a very sensible point prediction for the future observation,
$\hat u=e^{\hat v}=\hat\lambda=\bar y_n$.

Furthermore, the expected Hessian matrix is
%
\begin{eqnarray}\label{eq:fisherv1}
\hspace*{15pt}I_h(\lambda)&=& E_\lambda\left[-\pmatrix{
\dfrac{n+1}{\lambda^2}-2\dfrac{n \bar y_n +e^v}{\lambda^3} & \dfrac
{e^v}{\lambda^2}\vspace*{2pt}\cr
\dfrac{e^v}{\lambda^2} & -\dfrac{e^v}{\lambda}}\right]\hspace*{-15pt}\nonumber\\[-8pt]\\[-8pt]
&=&
\pmatrix{
\dfrac{n+1}{\lambda^2}& -\dfrac{1}{\lambda}\vspace*{2pt}\cr
-\dfrac{1}{\lambda} & 1}.\nonumber
\end{eqnarray}
It is easy to see that when evaluated at MLE\break ($=$MHLE), $\hat\lambda$,
$I_h(\hat\lambda)$
is identical to the \textit{observed Hessian matrix} $I_{h}^{\mathrm{obs}}=-\frac
{\partial^2 h(\phi;y)}{\partial\phi^2}|_{\phi=\hat\phi}$
%
\begin{eqnarray}
I_{h}^{\mathrm{obs}}&\equiv&-\pmatrix{
\dfrac{n+1}{\hat\lambda^2}-2\dfrac{n \bar y_n +e^{\hat v}}{\hat
\lambda^3} & \dfrac{e^{\hat v}}{\hat\lambda^2}\vspace*{2pt}\cr
\dfrac{e^{\hat v}}{\hat\lambda^2} & -\dfrac{e^{\hat
v}}{\lambda}}\nonumber\\[-8pt]\\[-8pt]
&=&
\pmatrix{
\dfrac{n+1}{\hat\lambda^2}& -\dfrac{1}{\hat\lambda}\vspace*{2pt} \cr
-\dfrac{1}{\hat\lambda} & 1},\nonumber
\end{eqnarray}
where the equality holds because $\hat\lambda=\bar y_n=e^{\hat v}$.
The fact that these two Hessian matrices coincide also gives us another
indication that the MHLE/Hessian matrix can behave just like MLE/Fisher
information for regular exponential families.

\subsection{So How Good is the Approximation?}
Now let us examine the inverse of $I_h(\lambda)$,
%
\begin{equation}\label{eq:fisherv2}
\qquad I_{h}^{-1}(\lambda)=
\pmatrix{
\dfrac{\lambda^2}{n}& \dfrac{\lambda}{n}\vspace*{2pt}\cr
\dfrac{\lambda}{n} & 1+\dfrac{1}{n}}\equiv\pmatrix{
\tau^2_\lambda& \tau_{\lambda, v} \vspace*{2pt}\cr
\tau_{\lambda, v} & \tau^2_v}
.
\end{equation}
If the $R$ term in (\ref{eq:taylor}) is negligible, then the above
matrix should
provide the (asymptotic) value of\break  \mbox{$V_\lambda(\hat\phi- \phi)$}
where $\phi=\{\lambda, v\}$
and the variance operator $V_\lambda$ is with respect to the \textit
{joint sampling distribution} $f_\lambda(y,v)$. Clearly, $\tau
_\lambda^2=\lambda^2/n$ is exactly right because it is $V_\lambda
(\hat\lambda)$. To examine the other entries, we first recall that
for \textit{large} $n$, Taylor's expansion (i.e., the $\delta$-method)
justifies the approximation
%
\begin{equation}\label{eq:delta}
\log(\bar y_n) - \log(\lambda) \approx\frac{\bar y_n - \lambda
}{\lambda} \equiv z_n.
\end{equation}
Adopting this approximation, and noting that $v=\log(y_{n+1})$ is
independent of
$\bar y_n$ given $\lambda$, we have
%
\begin{eqnarray}\label{eq:cov}
\operatorname{Cov}_\lambda(\hat\lambda, \hat v- v)&=&\operatorname
{Cov}_\lambda(\bar y_n, \log
(\bar y_n))\nonumber\\[-8pt]\\[-8pt]
&\approx&\operatorname{Cov}_\lambda(\bar y_n, z_n) = \frac
{\lambda}{n}\nonumber
\end{eqnarray}
which is the same as $\tau_{\lambda, v}$.

Similarly, by (\ref{eq:delta}), $V_\lambda(\log(\bar y_n))\approx
V(z_n)=1/n$, and hence we have
%
\begin{eqnarray}\label{eq:varv}
\qquad V_\lambda(\hat v- v)&=&V_\lambda(\log(\bar y_n))+V_\lambda(\log
(y_{n+1}))\nonumber\\[-8pt]\\[-8pt]
&\approx&
\frac{1}{n}+ V_\lambda(\log(y_{n+1})).\nonumber
\end{eqnarray}
This would be the same as $\tau^2_v$ if $V_\lambda(\log
(y_{n+1}))=1$. But unfortunately this is where the MHLE/Hessian matrix
approximation breaks down. One can directly verify or use the property
of Gumbel distribution (recall log of an exponential variable is a
Gumbel variable) to arrive at
%
\begin{equation}\label{eq:varl}
V_\lambda(\log(y_{n+1})) = \frac{\pi^2}{6} = 1.6449\!\ldots
\end{equation}
which is considerably larger than 1. [Incidentally, the integrating
moment generating function approach (Meng, \citeyear{meng2005}) can be used to
calculate $V_\lambda(\log(\bar y_n))$ exactly for general $n$, if
needed.]

%
\subsection{So What Works and What Does Not?}
To see more clearly what went wrong, let us write out the $R$ term in
(\ref{eq:taylor})
explicitly for the current model. Using
(\ref{eq:expov22}) and (\ref{eq:fisherv2}), simple algebra reveals
that (\ref{eq:taylor}) becomes
%
\begin{eqnarray}\label{eq:eees}
\pmatrix{
\hat\lambda-\lambda\vspace*{2pt}
\cr
\hat v - v}
&\equiv&
\pmatrix{
\bar y_n - \lambda\vspace*{2pt}
\cr
\log(\bar y_n) - \log(y_{n+1})}\nonumber\\[-8pt]\\[-8pt]
&=&
\pmatrix{
\bar y_n - \lambda\vspace*{2pt}
\cr
\dfrac{\bar y_n - y_{n+1}}{\lambda}}
+
\pmatrix{
0 \vspace*{2pt}
\cr
R_{v,n}},\nonumber
\end{eqnarray}
where $R_{v,n}$ obviously makes up the difference between $\hat v - v$
and $(\bar y_n - y_{n+1})/\lambda$, but it would be more useful to
express it in the equivalent form
%
\begin{eqnarray}\label{eq:rterm}
R_{v,n} &=& \biggl[\log\biggl(\frac{\bar y_n}{\lambda}\biggr)- \frac
{\bar y_n - \lambda}{\lambda}\biggr]\nonumber\\[-8pt]\\[-8pt]
&&{} - \biggl[\log\biggl(\frac
{y_{n+1}}{\lambda}\biggr)- \frac{y_{n+1} - \lambda}{\lambda}\biggr].\nonumber
\end{eqnarray}

From these expressions, we see that the MHLE/\break Hessian matrix approach
works perfectly for the estimation of $\lambda$---it is the same as
MLE and with the correct variance estimator because its $R$ term is
exactly zero. However, for the prediction of $v$, two things went
wrong, and both are due to the failure of accumulation of information.
First, $R_{v,n}$ is not negligible compared with the leading term
$Z_{v,n}=(\bar y_n - y_{n+1})/\lambda$. Indeed, as $n\rightarrow
\infty$, $R_{v,n} \rightarrow R_{\infty}=\xi- 1 - \log(\xi)$ and
$Z_{v,n}\rightarrow Z_\infty=1 - \xi$ where $\xi$ is an exponential
variable with mean one. In fact, while $E(Z_\infty)=0$,
$E(R_\infty)$ is far from zero, taking the value of Euler's constant,
$\gamma= 0.5772\ldots.$ This failure obviously is due to the
nonapplicability of the Taylor expansion (\ref{eq:delta}) \textit{when
$n=1$}; if this \textit{were} applicable, then $V(\log(y_{n+1}))=V
(v)$ would be approximated by $V(z_n)=1$, leading to $\tau
^2_v=1+\frac{1}{n}$ for $V(\hat v -v)$ in (\ref{eq:fisherv2}).

Second, although $Z_{\infty}$ has mean zero and variance one, its
density function $f(z)=e^{z-1}$, with support $(-\infty, 1]$,
is far from that of the normal. Indeed, $f(1)/\phi(1)>5$, where $\phi
(z)$ is the p.d.f. of $N(0,1)$. But of course the distribution of
$Z_\infty$ or $Z_{v, n}$ is not even relevant because we cannot use
either of them to approximate the sampling distribution of $\hat v -v$
due to the nonnegligibility of $R_{v,n}$.

%
\subsection{3-in-1: Pivotal Predictive Distribution, Posterior
Predictive Distribution, and H-distribution}
The~exact distribution of $\hat v -v$, of course, can be worked out
easily in this case.
But it is important to emphasize that by moving from the original
$u=y_{n+1}$ scale to the $v=\log(y_{n+1})$ scale, we have obtained a
\textit{predictive pivotal quantity}. That is, whereas the sampling
distribution of $u - \hat u =
y_{n+1} - \bar y_n$ depends on the unknown $\lambda$, the distribution
of $v - \hat v =
\log(y_{n+1}/\bar y_n)$ is free of $\lambda$ because it is canceled
in the ratio as the scale parameter. Consequently, the $v$ scale
provides us a way to construct exact prediction intervals without
having to worry about $\lambda$ which is a \textit{nuisance parameter
for the purposes of prediction.} This is simply the predictive version
of the usual inference of parameter of interest based on a pivotal
quantity. Although such a construction is by nature a frequentist one,
it should help to understand the importance of the choice of scale of
the unobservables for the authors' approach. Evidently, this
consideration of pivotal quantity greatly restricts the family of
scales for unobservables, beyond the minimal requirement of preserving
the (first two) Bartlett identities, as discussed in Section~\ref{sec5}.

Indeed, it is informative to compare the three distributions here: (I)
the sampling distribution $f_\lambda(\hat v - v)$, (II) the posterior
predictive distribution $f^B(v|y)$ under constant prior and (III) the
h-distribution $f^H(v|y)$ derived from the authors' APHL method. For
(I), because $U_n=\sum_{i=1}^n y_i \sim \operatorname{Gamma}(n, \lambda)$ is
independent of $u=y_{n+1} \sim \operatorname{Gamma}(1, \lambda)$, we know the ratio
$B_n=U_n/(U_n+u)$ is distributed as $\operatorname{Beta}(n, 1)$. Consequently,
$r=y_{n+1}/\bar y_n=n(B^{-1}_n-1)$ follows a Pareto distribution of
order $n+1$, that is,
%
\begin{equation}\label{eq:pareto}
f(r) = \biggl(1 + \frac{r}{n}\biggr)^{-(n+1)},\quad    r \ge0
\end{equation}
which converges to $e^{-r}$ as $n\rightarrow\infty$, as it should.
[The~distribution $f(r)$ obviously determines the distribution of
$v-\hat v=\log(r)$.]

In comparison, for (II), because $f(y_1,\ldots, y_n|\lambda) \propto
\lambda^{-n}e^{-U_n/\lambda}$, a posteriori we can write
$\lambda= U_n \gamma^{-1}$, where $\gamma\sim \operatorname{Gamma}(n-1, 1)$.
Consequently, because $u=\lambda\xi$ where $\xi\sim \operatorname{Gamma}(1,1)$ and
is independent of $\gamma$, a posteriori we have $u=U_n (\xi
/\gamma)$. This implies $r\equiv nu/U_n=n\xi/\gamma=n(\tilde B_{n-1}-1)$
where $\tilde B_{n-1} \sim\break \operatorname{Beta}(n-1, 1)$; here we assume $n>1$ as the
posterior is improper when $n=1$ under the constant prior on $\lambda
$. It follows that
%
\begin{equation}\label{eq:ppareto}
f^B(r|y) = \frac{n-1}{n}\biggl(1 + \frac{r}{n}\biggr)^{-n},\quad    r
\ge0.
\end{equation}

For (III), we note from the first equation of (\ref{eq:expov22}) that
for any given $v$,
the h-likelihood is maximized at
%
\begin{equation}\label{eq:profl}
\lambda(v) = \frac{n\bar y_n + e^v}{n+1}.
\end{equation}
From (\ref{eq:expov1}), the log profile h-likelihood then becomes,
ignoring irrelevant
constants,
%
\begin{equation}\label{eq:profh}
h_\lambda(v; y)= -(n+1)\log\lambda(v)+v.
\end{equation}
Using the authors' notation and (\ref{eq:fisherv1}), $D(h,\lambda
)=-\frac{\partial^2 h(\lambda, v:y)}{\partial\lambda
^2}=(n+1)/\lambda^2$ when $\lambda=\lambda(v)$, and hence the
authors' (log) adjusted
profile h-likelihood becomes, again ignoring irrelevant constants,
%
\begin{eqnarray}\label{eq:aphl}
\tilde h_\lambda(v; y)&=& -(n+1)\log\lambda(v)+v\nonumber\\
&&{}- \tfrac{1}{2}\log
(D(h,\lambda(v)))\\
&=&
-n\log\lambda(v)+v.\nonumber
\end{eqnarray}
The~h-distribution for $v$ then, as I understand from the authors'
approach, is to set
%
\begin{eqnarray}\label{eq:aphd}
f^H(v|y)&\propto& e^{\tilde h_\lambda(v; y)}\nonumber\\[-8pt]\\[-8pt]
& =& e^{v}\lambda
^{-n}(v)\propto e^{v}(U_n+e^v)^{-n}.\nonumber
\end{eqnarray}
Converting this to the distribution of $r=nu/U_n=ne^{v}/U_n$ and
re-normalizing it to be a proper distribution, we have, again assuming $n>1$,
%
\begin{equation}\label{eq:hpareto}
f^H(r|y) = \frac{n-1}{n}\biggl(1 + \frac{r}{n}\biggr)^{-n}, \quad   r
\ge0
\end{equation}
which is identical to the posterior predictive distribution (\ref
{eq:ppareto}). This is expected because of the accuracy of the Laplace
approximation (and by re-normalizing we
eliminate the remaining approximation inaccuracy).

%
\subsection{The~Need of Choosing the Right Scale for the Fixed Parameter}

A perceptive reader may realize that the small difference between (\ref
{eq:pareto}) and (\ref{eq:ppareto}) or (\ref{eq:hpareto}), although
of little practical consequence, nevertheless points to a deeper issue.
Indeed, if we use the constant prior on $\log(\lambda)$, the most
common ``noninformative'' prior for scale parameter, then $f^B(r|y)$
will be the same as $f(r)$ of (\ref{eq:pareto}). This suggests an
intimate connection between posterior prediction and the pivotal
approach on the joint space of $\{y, v\}$.

For h-likelihood, we have seen that choosing the right scale for the
unobservable is crucial. However, the scale of the parameter also plays
a role, especially for the adjusted profile h-likelihood because the
value of $D(h,\alpha)$ depends on the scale of $\alpha$. For example,
in the current example, if we also choose the log scale for $\lambda$,
that is,
use $h(\eta, v;y)$ to carry out all the h-likelihood calculations
where $\eta=\log(\lambda)$,
then $D(h,\eta)=n+1$. Consequently, the adjustment becomes immaterial,
making the log APHL the same as (\ref{eq:profh}), the original
profiled log h-likelihood.
This is easily seen to lead to
%
\begin{equation}\label{eq:hpare}
f^H(r|y) = \biggl(1 + \frac{r}{n}\biggr)^{-(n+1)},\quad    r \ge0,
\end{equation}
which is now identical to the pivotal predictive distribution $f(r)$ in
(\ref{eq:pareto}), a truly 3-in-1!

This equivalence not only demonstrates the intimate connection among
the three methods,
but also suggest the possibility of providing a probabilistic meaning to
h-distributions, at least in some cases. For example, under (\ref
{eq:pareto}), a $1-\alpha$ \textit{highest density predictive} (HDP)
interval is of the form
%
\begin{eqnarray}\label{eq:hdp}
\mathit{HDP}=[0, \ c(\alpha, n) \bar y_n],\\
  \eqntext{\mbox{where }   c(\alpha,
n)= n(\alpha^{-{1}/{n}} -1)\rightarrow-\log(\alpha).}
\end{eqnarray}
This interval has both Bayesian interpretation and frequentist
interpretation, the latter of which I believe is closer to what the
authors have been seeking. The~frequentist interpretation is simply
that among repeated samples of $\{y_1,\ldots, y_n, y_{n+1}\}$, the HDP
in (\ref{eq:hdp}) covers $y_{n+1}$ with frequency/probability
$1-\alpha$. Such interpretation perhaps is more appealing to some than
its posterior predictive interpretation which in this case is actually
not directly realizable with random $\lambda$ because it is derived
under the improper prior $\pi(\lambda)\propto\lambda^{-1}$. It is
somewhat intriguing that this un-realizable posterior predictive
distribution via random $\lambda$ is easily realizable via the pivotal
predictive distribution. A general investigation of this connection may
offer new insights into both the similarities and differences between
Bayesian and sampling inferences.

\section{Epilogue}
Dan Brown concluded \textit{Angels and Demons}
with Dr.~Langdon's religious experience with Vittoria,\break  a~yoga master.
Although my pleasure is at an entirely different level, I must confess
that my study of the \mbox{h-likelihood} framework is largely carried by both
the authors' faith in their methods and my faith in the authors---they
must have seen signs that most discussants did not. My Bayesian half
urged me every weekend to
seek Dr. Langdon's ambigram of ``H,'' yet my other half kept seducing me
with promises of hidden treasures. Indeed, a posteriori I am
willing to move all probability from (V) to (IV), as well as to
increase the probability of (II) over 50\%, provided that we are always
mindful of another ``H'' for \mbox{h-likelihood}---its Achilles' Heel---the
potential (and often) non-negligibility of the $R$ term. The~Bartlizability and pivotal predictive interpretation of the
\mbox{h-likelihood} methods could seduce someone to speculate that the ``H'' is
\textit{The~Lost Symbol}, the eagerly awaited new thriller of Dan Brown.
As a matter of fact, since I have already been seduced for the past
five weekends, far exceeding the originally planned 3-day excursion,
I may as well enjoy my earned fantasy, a spoonful of my colleague Dr.
Langdon's new experience, divine or not\ldots.

\section*{Acknowledgments}
I thank Professors Lee and Nelder for inspiring me over five rewarding
weekends, and possibly many more. I also thank NSF for partial funding,
and Joe Blitzstein, Yves Chretien and Xianchao Xie for their
proofreading and very helpful comments. Any hallucination, of course,
is mine.

\vspace*{-2pt}
\end{document}